\documentclass[aps,pre,showpacs,
amssymb,twocolumn
]{revtex4-2}
\usepackage{graphicx}
\usepackage{amsmath}
\usepackage{amsfonts}
\usepackage{amssymb}
\usepackage{epsfig,libertine}
\usepackage[libertine]{newtxmath}
\usepackage{color}
\usepackage{dsfont, hyperref, xcolor}
\usepackage{comment}
\usepackage{enumerate}

\newcommand{\abs}[1]{\left\vert#1\right\vert}

\newcommand{\Tr}[1]{\text{Tr}\left\{#1\right\}}

\newcommand{\bra}[1]{\langle#1\vert}
\newcommand{\ket}[1]{\vert#1\rangle}

\begin{document}

\title{Fluctuation theorems and expected utility hypothesis}

\author{Gianluca~Francica, Luca Dell'Anna}
\address{Dipartimento di Fisica e Astronomia e Sezione INFN, Universit\`{a} di Padova, via Marzolo 8, 35131 Padova, Italy}

\date{\today}

\begin{abstract}
The expected utility hypothesis is a popular concept in economics that is useful for making decisions when the payoff is uncertain.
In this paper, we investigate the implications of a fluctuation theorem in the theory of expected utility. In particular, we wonder whether entropy could serve as a guideline for gambling. We prove the existence of a bound involving the certainty equivalent which depends on the entropy produced. Then, we examine the dependence of the certainty equivalent on the entropy by looking at specific situations, in particular the work extraction from a nonequilibrium initial state.
\end{abstract}

\maketitle

\section{Introduction}
Wealth is commonly affected by uncertainty, which can come from our impossibility to know all the information to get a deterministic prediction. For instance, in a quantum system this uncertainty can be intrinsic and cannot be reduced arbitrarily.
Thus, typically an agent aims to gain a wealth
which is affected by stochastic fluctuations, so that he has to make a choice (among the different procedures available) under uncertainty. A risk neutral agent will make his choice by preferring procedures giving the maximum average gain. However, if the agent takes in account the risk of his choice, he can make his choice relying on the expected utility hypothesis, first formalized by von Neumann and Morgenstern within the theory of games and economic behaviour in 1944~\cite{vonNeumann}. Thus, the choice will depend on how much the agent is risk averse. Formally, risk aversion can be characterized in terms of a utility function, from which we get the certainty equivalent (in simple terms, the certain amount of wealth that is equivalent, from the agent's point of view, to the procedure in question). Fluctuations are not always arbitrary, e.g., the fluctuations coming from the permanent state of thermal agitation of the matter cannot violate the second law of thermodynamics. This behavior can be a consequence of the so-called fluctuation theorems (see, e.g., Refs.~\cite{jarzynski97,crooks99,campisi11}).

From a thermodynamic point of view, the role of wealth can be related to the thermodynamic work (see, e.g., Ref.~\cite{Ducuara23}), which is a stochastic quantity. When the initial state is an equilibrium Gibbs state, the work $w$ satisfies the Crooks fluctuation theorem
\begin{equation}\label{eq.crooks}
p(w) = e^{\beta(w-\Delta F)}\bar{p}(-w)\,,
\end{equation}
which relates the work probability distribution $p(w)$ of the out-of-equilibrium process to the probability distribution $\bar{p}(w)$ of a corresponding time-reversed process. In particular, $\Delta F$ is the equilibrium free energy change, so that $w_{irr}=w-\Delta F$ gives the irreversible work. While an agent that is neutral to risk will prefer to minimize the average irreversible work $\langle w_{irr}\rangle$ in the out-of-equilibrium process, in general a risk non-neutral agent will look on the irreversible work certainty equivalent, denoted with $w^{CE}_{irr}$, which for a constant absolute risk aversion $r=\alpha-1$, can be expressed as $w^{CE}_{irr} = \beta^{-1} D_\alpha(p(w)||\bar{p}(-w))$, where $D_\alpha(p(w)||\bar{p}(-w))$ is the R\'{e}nyi divergence (see Ref.~\cite{Ducuara23} for details). 
We note that $w$ can take both positive and negative values, although in microeconomics the utility is generally defined just for positive values of the variables.
The result of Ref.~\cite{Ducuara23} suggests that the fluctuation theorems can play a role in the expected utility hypothesis, and here we will try to explore this role further. Differently from Ref.~\cite{Ducuara23}, we aim to derive some general results which can be applied to the work extraction from a nonequilibrium initial state, so that the fluctuation theorem in Eq.~\eqref{eq.crooks} does not hold. Our idea is to look on the work extracted certainty equivalent instead of the irreversible work one as done in Ref.~\cite{Ducuara23}. This leads to a `game' different from the one introduced in Ref.~\cite{Ducuara23}. To do this, we will start from the thermodynamic uncertainty relations of Refs.~\cite{Hasegawa19,Timpanaro19,Francica22}, which provide a bound of the fluctuations of some random quantity. The bound follows from some fluctuation theorems involving the entropy produced and the quantity, as the fluctuation theorem of Ref.~\cite{Garcia10}.
We prove a generalized thermodynamic uncertainty relation, which involves the certainty equivalent for a risk non-neutral agent and reproduces the usual thermodynamic uncertainty relation when the agent is risk neutral.
This result shows how the certainty equivalent is related to the average entropy produced. However, there are situations where the certainty equivalent shows a non-trivial dependence on the entropy, which goes beyond this bound.


\section{Expected utility hypothesis and fluctuation theorems}
For our purposes, we focus on an agent who must choose between two procedures that yield two different wealths represented by the random variables $w_1$ and $w_2$ (having certain probability distributions).
To give an example, we consider an agent who must choose between winning a fixed payoff $w_{det}=50$ or flipping a coin and winning a payoff $w_{head}=100$ if heads or nothing otherwise. Intuitively, if the agent is risk neutral, he is indifferent to the choice, since if he flips the coin he will win the average payoff $w_{det}$. Furthermore, we expect that if the agent is risk averse (loving), he will choose the certain payoff $w_{det}$ (of flipping the coin).
The expected utility theory modelizes the agent's risk aversion by using a utility function $u(w)$, so that (following this theory) the agent will choose the procedure yielding the wealth $w_1$ instead of $w_2$ if~\cite{bookmicroeco,bookmicroeco2}
\begin{equation}\label{eq.exp utility theo}
\langle u(w_1) \rangle > \langle u(w_2) \rangle\,.
\end{equation}
Thus, in our example, an agent with the utility function $u(w)$ will choose to flip the coin if  $u(0)+u(w_{head})> 2 u(w_{det})$, is neutral to the choice if equality holds or will choose certain payoff otherwise.
It is easy to see that the inequality in Eq.~\eqref{eq.exp utility theo} remains unchanged if we perform an affine transformation on the utility function, i.e., the transformation $u(w) \mapsto a u(w) + b$, where $a$ is a positive variable. This means that the utility function is defined up to affine transformations, since two utility functions related by such transformation gives the same preference ordering given by Eq.~\eqref{eq.exp utility theo}.
The wealth $w$ can be further characterized by the certainty equivalent, denoted with $w_{CE}$, defined such that
\begin{equation}\label{eq.wCEdef}
u(w_{CE}) = \langle u(w) \rangle\,.
\end{equation}
Thus, the certainty equivalent is obtained as the Kolmogorov-Nagumo average of the wealth, i.e., $w_{CE}=u^{-1}(\langle u(w)\rangle)$.
The meaning of the certainty equivalent becomes more clear if we consider as usual a strictly increasing utility function $u(w)$, so that Eq.~\eqref{eq.exp utility theo} is equivalent to $w^{CE}_1> w^{CE}_2$, where $w^{CE}_{1,2}$ is the certainty equivalent corresponding to the wealth $w_{1,2}$.
To understand in simple terms how the agent's risk aversion depends on the utility function $u(w)$, we start by noting that if $u(w)$ is a linear function the certainty equivalent coincides with the average value, i.e., $w_{CE}=\langle w \rangle$. In this case, the agent prefers the procedure maximizing the average wealth and it is neutral to risk. Instead, if $u(w)$ is a strictly increasing concave function then the agent is averse to risk, since by applying the Jensen's inequality to Eq.~\eqref{eq.wCEdef} we get $w_{CE}<\langle w \rangle$. Similarly, if $u(w)$ is a strictly increasing convex function the agent will be loving to risk since $w_{CE}>\langle w \rangle$. In particular, the difference $RP=\langle w \rangle - w_{CE} $ gives the so-called risk premium.
For instance, in our example, if $u(w)$ is a convex function (for arbitrary values $w_{head}$ and $w_{det}$), by applying Jensen's theorem we get the inequality $u(0) + u(w_{head})\geq 2 u(w_{head}/2)$.
Then, by considering $u(w)$ strictly increasing, if $w_{head} > 2 w_{det}$ from the previous discussion the agent will flip the coin.
In summary, we say that the agent is risk averse if $w_{CE}<\langle w\rangle$, risk neutral if $w_{CE}=\langle w\rangle$ and risk loving if $w_{CE}> \langle w\rangle$.
Furthermore, the utility function allows us to quantify how risk averse an agent is. For a utility function which is concave and strictly increasing, risk aversion can be measured with the Arrow-Pratt coefficient of absolute risk aversion defined as
\begin{equation}\label{eq. RA}
r_A(w) = -\frac{u''(w)}{u'(w)}\,,
\end{equation}
which is non-negative. It is clear that risk aversion depends on how much the utility function $u(w)$ is concave. Then, the more simple quantifier of risk aversion should be the second derivative $u''(w)$. However, $u''(w)$ is not invariant under affine transformations, but we can work around this problem by dividing it by the first derivative $u'(w)$, which explains Eq.~\eqref{eq. RA}. More details can be found, e.g., in Refs.~\cite{bookmicroeco,bookmicroeco2}.

\subsection{Certainty equivalent bound}
Having introduced some rudiments of the expected utility hypothesis, we now focus on a different hypothesis that the wealth $w$ is related to a random variable $\sigma$ by the fluctuation theorem
\begin{equation}\label{eq.fluc}
\frac{p(\sigma,w)}{p(-\sigma,-w)} = e^\sigma\,,
\end{equation}
where $p(\sigma,w)$ is the joint probability distribution of the variable $\sigma$ and the wealth $w$. For instance this theorem holds
for specific physical variables $w$ and $\sigma$, e.g., see Ref.~\cite{Garcia10}.
So, making an analogy with thermodynamics, we will call $\sigma$ stochastic entropy. Thus, we can assume that the fluctuation theorem in Eq.~\eqref{eq.fluc} holds as happens in certain physical situations (e.g., the ones of the thermodynamic uncertainty relations of Refs.~\cite{Hasegawa19,Timpanaro19}), and that the quantity $w$ can be thought of as wealth.
For our purposes the variable $w$ will be thought of as the work extracted in Sec.~\ref{sec. work extraction}, so that the fluctuation theorem can hold for some initial nonequilibrium states, as we will explicitly show.
In general, given a wealth $w$ with probability distribution $p(w)$, there can exist an entropy $\sigma$ such that the joint probability distribution $p(\sigma,w)$ satisfies the fluctuation theorem in Eq.~\eqref{eq.fluc}. For instance, when the support of $p(w)$ is discrete, a sufficient condition for the existence of a variable $\sigma$ such that Eq.~\eqref{eq.fluc} holds, is that the probability distribution of the wealth has the form $p(w)=\sum_{n=-N}^N p_n \delta(w-w_n)$, with $w_n=-w_{-n}$ and $p_n p_{-n}= K^2$, where $K$ is some constant. In this case, we can define $p(\sigma,w) =\sum_{n=-N}^N p_n \delta(\sigma - \ln(p_n/p_{-n}))\delta(w-w_n)$ and we can easily check that Eq.~\eqref{eq.fluc} is satisfied. In particular,  $p(w)$ is the marginal probability distribution $p(w)=\int p(\sigma,w)d\sigma$. However, we can also make the trivial choice $p(\sigma,w)=\delta(\sigma) p(w)$, so that the joint probability distribution $p(\sigma,w)$ resulting from $p(w)$ is not unique. In general, for discrete supports of $w$ and $\sigma$, which are $\{w_n\}$ and $\{a_k\}$, we need to have $w_{n}=-w_{-n}$ and $a_k=-a_{-k}$, i.e., the joint probability distribution needs to have the form $p(\sigma,w) =\sum_{k=-M}^M\sum_{n=-N}^N p_{k,n} \delta(\sigma - a_k)\delta(w-w_n)$, with $M$ arbitrary integer, where $\sum_{k,n}p_{k,n}=1$ and $p_{k,n}\geq 0$. From Eq.~\eqref{eq.fluc} the integral $p(-w)= \int e^{-\sigma}p(\sigma,w)d\sigma$  gives a condition for the probabilities $p_{n}=\sum_k p_{k,n}$ of the wealth, depending on $M$ and typically different from the condition $p_n p_{-n}= K^2$, which is achieved for $M=N$.
In general, a necessary condition for the existence of $\sigma$ such that Eq.~\eqref{eq.fluc} holds, is that $p(w)$ has the same support of $p(-w)$, since Eq.~\eqref{eq.fluc} implies that $p(-w)= \int e^{-\sigma}p(\sigma,w)d\sigma$ and the integration over $\sigma$ changes the probabilities but does not change the support of $w$.  Then, given a  wealth $w$, it is not always possible to define $p(\sigma,w)$ such that Eq.~\eqref{eq.fluc} holds: Eq.~\eqref{eq.fluc} is an hypothesis whose implications to the expected utility we want to examine.
To be as general as possible, we consider an agent with an arbitrary utility function $u(w)$, which  in general can always be written as $u(w)=u_e(w)+u_o(w)$, with $u_e(w)=u_e(-w)$ and $u_o(w)=-u_o(-w)$. In particular, given any utility function $u(w)$, the even and odd components can be obtained as $u_{e,o}(w)=(u(w)\pm u(-w))/2$.
If the fluctuation theorem in Eq.~\eqref{eq.fluc} holds, we get the general bound
\begin{equation}\label{eq.main}
\frac{(u(w_{CE})-\langle u_e(w)\rangle)^2}{\langle u_o^2(w)\rangle}\leq  \left\langle \tanh^2\left(\frac{\sigma}{2}\right) \right\rangle\leq f^2(\langle \sigma\rangle)\,,
\end{equation}
where $f$ is the inverse of $h(x)=2x\tanh^{-1} x$. Thus, the certainty equivalent is constrained by the entropy production.
Proof: To prove Eq.~\eqref{eq.main}, we note that
\begin{equation}\label{eq.1}
u(w_{CE}) = \langle u_e(w) \rangle + \langle u_o(w) \rangle\,,
\end{equation}
thus we get
\begin{equation}\label{eq.trivial}
(u(w_{CE})-\langle u_e(w)\rangle)^2= \langle u_o(w)\rangle^2\leq \langle u_o^2(w)\rangle\,,
\end{equation}
which is a trivial bound which does not involve explicitly the entropy.
From the fluctuation theorem in Eq.~\eqref{eq.fluc}, given an arbitrary function of two-variables $F(\sigma,w)$ we get the identity
\begin{equation}\label{eq.id}
\langle F(\sigma,w) \rangle = \langle F(-\sigma,-w)e^{-\sigma}\rangle\,,
\end{equation}
from which
\begin{eqnarray}
\langle u_o(w) \rangle &=& \frac{1}{2} \langle u_o(w) (1-e^{-\sigma})\rangle\\
 &=& \frac{1}{2} \langle u_o(w)\sqrt{1+e^{-\sigma}} \frac{(1-e^{-\sigma})}{\sqrt{1+e^{-\sigma}}}\rangle\,.
\end{eqnarray}
By using the Cauchy-Schwartz inequality we get
\begin{eqnarray}
\langle u_o(w) \rangle^2 &\leq& \frac{1}{4} \left\langle u_o^2(w)(1+e^{-\sigma})\right\rangle \left\langle \frac{(1-e^{-\sigma})^2}{1+e^{-\sigma}}\right\rangle\\
 &=& \langle u_o^2(w)\rangle \left\langle \tanh^2\left(\frac{\sigma}{2}\right) \right\rangle\,,
\end{eqnarray}
where we have used the identity in Eq.~\eqref{eq.id}.
As shown, e.g., in Ref.~\cite{Francica22}, we have the inequality
\begin{equation}
\left\langle \tanh^2\left(\frac{\sigma}{2}\right) \right\rangle  \leq f^2(\langle \sigma\rangle )\,,
\end{equation}
from which follows Eq.~\eqref{eq.main}. $\square$

The bound achieved is tighter than the trivial one obtained from Eq.~\eqref{eq.trivial} since $0\leq f^2(x)\leq \tanh(x/2)\leq 1$, with $f(0)=0$ and $f(x)\to 1$ as $x\to \infty$ (see, e.g., Ref.~\cite{Francica22}).
In particular, from Eq.~\eqref{eq.main} we see that as $\langle \sigma \rangle \to 0$, if $\langle u_o^2(w)\rangle \neq 0$ and it is finite, then $w_{CE}\to u^{-1}( \langle u_e(w)\rangle)$. E.g., if $p(\sigma,w)=\delta(\sigma) p(w)$, from the fluctuation theorem we need to have $p(w)=p(-w)$,  and this result directly follows from Eq.~\eqref{eq.1}.
If $\langle \sigma \rangle \to \infty$, from the bound we get Eq.~\eqref{eq.trivial} so that the entropy tends to do not constrain the certainty equivalent.
The bound can be saturated, i.e., following Ref.~\cite{Timpanaro19}, we consider the minimal probability distribution
distribution $p_{min}(\sigma,w)$, which reads
\begin{eqnarray}\label{eq.minimal}
\nonumber p_{min}(\sigma,w) &=& \frac{1}{2\cosh(a/2)}\big( e^{a/2}\delta(\sigma-a)\delta(w-b)\\
 && + e^{-a/2}\delta(\sigma+a)\delta(w+b)\big)\,.
\end{eqnarray}
 Then, we get the equality in Eq.~\eqref{eq.main} for $p(\sigma,w)=p_{min}(\sigma,w)$ or for $p(\sigma,w)=\delta(\sigma)p(w)$.
For a linear utility function, so that the certainty equivalent is $w_{CE}=\langle w \rangle$, Eq.~\eqref{eq.main} reduces to
\begin{equation}\label{eq.theunc}
\frac{w_{CE}^2}{\langle w^2\rangle}\leq f^2(\langle \sigma\rangle)\,,
\end{equation}
which is the thermodynamic uncertainty relation of Ref.~\cite{Timpanaro19}.
Thus, for a risk neutral agent, this bound suggests that if $\langle w \rangle >0$ ($\langle w \rangle <0$) the certainty equivalent becomes smaller (larger) as the entropy decreases, when $\langle w^2\rangle$ changes slowly with the entropy.
Equivalently, Eq.~\eqref{eq.main} can be written as
\begin{equation}\label{eq.b1}
\langle u_e(w) \rangle<u(w_{CE}) \leq \langle u_e(w) \rangle + f(\langle \sigma \rangle) \sqrt{\langle u_o^2(w)\rangle}
\end{equation}
if $\langle u_o(w)\rangle >0$, or as
\begin{equation}\label{eq.b2}
\langle u_e(w) \rangle>u(w_{CE}) \geq \langle u_e(w) \rangle - f(\langle \sigma \rangle) \sqrt{\langle u_o^2(w)\rangle}
\end{equation}
if $\langle u_o(w) \rangle <0$.

A relation between the certainty equivalent $w_{CE}$ and $\sigma$ can emerge from the bound. When the expectation values $\langle u_e(w) \rangle$ and $\langle u_o^2(w)\rangle$ do not change with $\langle \sigma \rangle$, the right side of Eq.~\eqref{eq.b1} decreases as the average entropy decreases, whereas the right side of Eq.~\eqref{eq.b2} increases as the average entropy decreases. This suggests that a risk non-neutral agent prefers procedures producing a small or large average entropy depending on the sign of  $\langle u_o(w) \rangle$, since the latter determines the validity of one of Eqs.~\eqref{eq.b1}-\eqref{eq.b2}.
However, outside of this case the relation between $w_{CE}$ and $\langle\sigma\rangle$ can be very complex. To see it, we note that for the distribution in Eq.~\eqref{eq.minimal} the bound is saturated and both $\langle u_e(w) \rangle$ and $\langle u_o^2(w)\rangle$ do not change with $\langle \sigma \rangle$, but if we add a point in the support we get
\begin{eqnarray}
\nonumber p_{3}(\sigma,w) &=& \frac{1}{1+2\cosh(a/2)}\big( e^{a/2}\delta(\sigma-a)\delta(w-b)\\
 && + e^{-a/2}\delta(\sigma+a)\delta(w+b)+\delta(\sigma)\delta(w)\big)\,,
\end{eqnarray}
so that clearly both $\langle u_e(w) \rangle$ and $\langle u_o^2(w)\rangle$ will change with $\langle \sigma \rangle$. They change slowly if $u_e(x)$ and $u_o(x)$ evaluated at $x=0,b$ are not too large, e.g., it is easy to see that $\langle u_e(w)\rangle$ monotonically changes from $(u_e(0)+2u_e(b))/3$ to $u_e(b)$ as  the average entropy goes from zero to infinity. Then, if $\langle u_o(w)\rangle >0$, so that Eq.~\eqref{eq.b1} holds, if $(u_e(0)+2u_e(b))/3\approx u_e(b)$, $u(w_{CE})$ will decrease as the average entropy decreases, but if $u_e(b)$ is negative and very large in modulus, $u(w_{CE})$ follows the trend of $\langle u_e(w)\rangle$ and it will increase as the average entropy decreases.

\subsection{Exponential utility function: examples}
\begin{figure}
[t!]
\centering
\includegraphics[width=0.87\columnwidth]{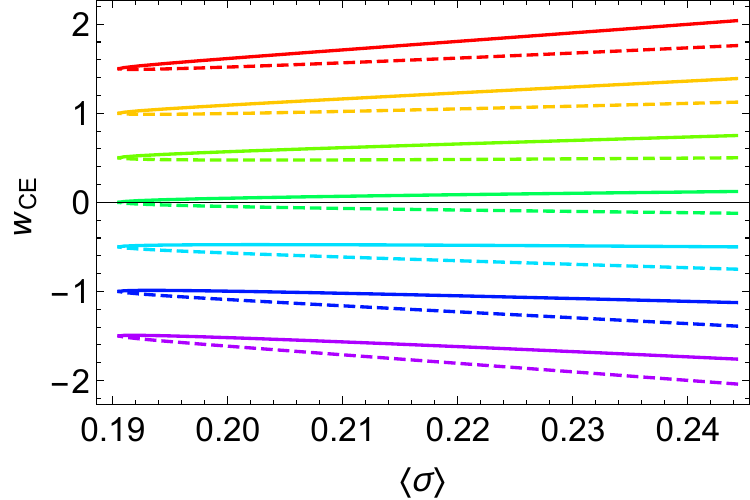}
\caption{ The parametric plot of $w_{CE}$ versus the entropy $\langle\sigma\rangle$ for different values of $r$. We change $\gamma$ from zero to one (solid lines) and from minus one to zero (dashed lines). We put $r=-6,-4,-2,0,2,4,6$ from top to bottom.  Of course, the curves keep the trend which emerges in the plot also for larger values of $r$.
}
\label{fig:plotfin1}
\end{figure}

\begin{figure}
[t!]
\centering
\includegraphics[width=0.47\columnwidth]{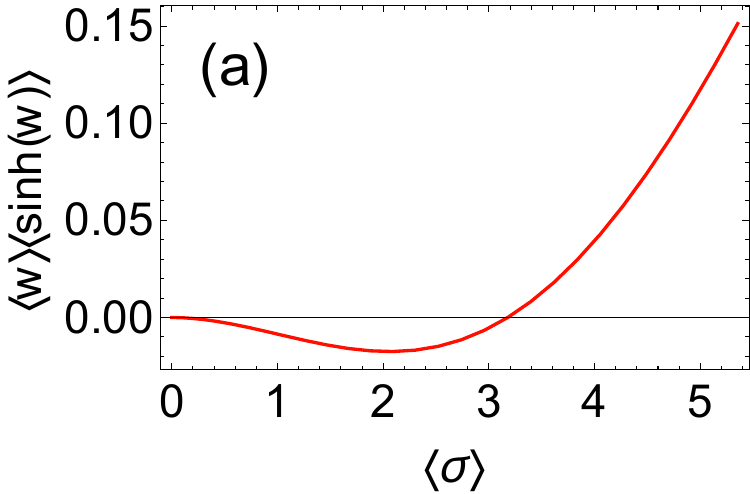}\includegraphics[width=0.47\columnwidth]{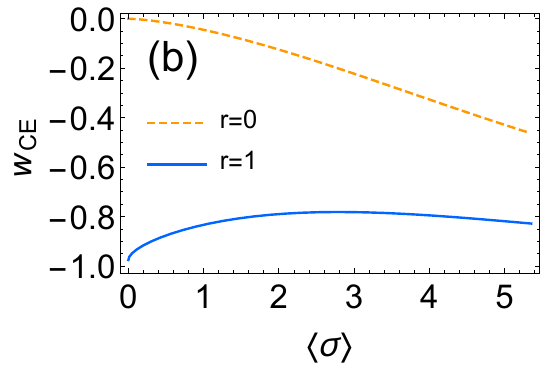}
\caption{ The parametric plots of $\langle w\rangle \langle \sinh (rw) \rangle $ (panel (a)) and the certainty equivalent $w_{CE}$ (panel (b)) versus the entropy $\langle\sigma\rangle$. We consider $N=2$, $w_n=n$ for $n>0$, $a_1=-2 a_2$. The plot is generated by changing $a_2$ in $[0,3]$.
}
\label{fig:plot0}
\end{figure}

To be more quantitative we give some mathematical examples by focusing on the particular case of the exponential utility function defined such that
\begin{equation}\label{eq.uteqxpo}
u(w) = \frac{1}{r}(1-e^{-rw})
\end{equation}
for $r\neq 0$, and $u(w)=w$ for $r=0$, which is a strictly increasing function. The agent is risk averse for $r>0$, risk neutral for $r=0$ and risk loving for $r<0$, and the absolute risk aversion of Eq.~\eqref{eq. RA} is constant and it is $r_A(x)=r$. From Eq.~\eqref{eq.wCEdef}, defining the certainty equivalent $w_{CE}$, we get
\begin{equation}
e^{-rw_{CE}} = \langle e^{-rw}\rangle\,.
\end{equation}
The even and odd components of the utility function $u(w)$ can be easily calculated as explained above and explicitly read $u_e(x)=(1-\cosh(r x))/r$ and $u_o(x)=\sinh(rx)/r$.
If $\langle \sinh r w \rangle >0$ when $r>0$, e.g., when $w$ takes non-negative values (we cannot have losses), Eq.~\eqref{eq.b1} is achieved which suggests that the certainty equivalent becomes smaller as the entropy decreases.
In contrast, if $\langle \sinh r w \rangle <0$ when $r>0$ (in simple terms losses are very likely), Eq.~\eqref{eq.b2} is achieved which suggests that the certainty equivalent becomes larger as the entropy decreases. However, for large $r$, since $\langle u_e(w) \rangle$ and/or $\langle u_o^2(w)\rangle$ can change strongly with the average entropy, we can obtain the opposite situation as we have seen for a distribution with a support of three points. To give another example, we consider the probability distribution
\begin{equation}
p(\sigma,w)= n (g(\sigma,w)\theta(\sigma)+g(-\sigma,-w)e^\sigma\theta(-\sigma))\,,
\end{equation}
where $n$ is such that $\int p(\sigma,w)d\sigma dw =1$ and $g(\sigma,w)$ is a non-negative function (see, e.g., Ref.~\cite{campisi21}). We focus on $g(\sigma,w)=e^{-\sigma^2-w^2+\gamma \sigma w}$. In this case, $\langle w\rangle$ has always the same sign of $\langle \sinh r w \rangle$ with $r>0$, which is equal to the sign of $\gamma$.
As shown in Fig.~\ref{fig:plotfin1}, for $\gamma>0$ ($\gamma<0$) the certainty equivalent $w_{CE}$ increases (decreases) as the entropy increases for $r$ non-positive (non-negative), whereas $w_{CE}$ decreases (increases) as the entropy increases for $r$ positive (negative) and large in modulus. Thus, every risk loving (averse) agent prefers to gamble when the entropy is large (small) when the average gain $\langle w\rangle$ is positive (negative). On the other hand, when $\langle w\rangle$ is negative (positive), this becomes true only as the agent becomes more and more risk loving (averse).
Furthermore, even the bound suggests that a risk non-neutral agent can prefer the opposite situation than a risk neutral one, e.g., we can have $\langle w\rangle <0$ and $\langle \sinh (rw) \rangle >0$ with $r>0$, so that the bound suggests that a risk neutral agent prefers to gamble when $\langle \sigma \rangle$ is small, whereas a risk non-neutral agent prefers $\langle \sigma \rangle $ large.
This can happen for a distribution with more than three points in the support, e.g., of the form
\begin{equation}
p_{2N}(\sigma,w)=K\sum_{n=-N,n\neq 0}^N e^{a_n/2} \delta(\sigma-a_n)\delta(w-w_n)
\end{equation}
with $K^{-1}=2 \sum_{n>0} \cosh(a_n/2)$, the support such that $w_n=-w_{-n}$ and $a_n=-a_{-n}$ to ensure the distribution satisfies the fluctuation theorem. As shown in Fig.~\ref{fig:plot0} for $N=2$, in the region where $\langle w\rangle <0$ and $\langle \sinh (rw) \rangle >0$ with $r>0$, the certainty equivalent increases when the entropy increases, as suggested by the bound. In particular, we note that for $r=1$, the  certainty equivalent is maximum for a non-zero $\langle \sigma \rangle$, corresponding to a negative $\langle w \rangle$, so that the agent prefers to gamble even if on average the losses are different from zero. Of course, this behavior will be achieved for more points in the support, e.g., we get it if the probability is peaked only on four points in the support, so that the examined case in Fig.~\ref{fig:plot0} is approximately reproduced.

\section{Work extraction}\label{sec. work extraction}
In order to show how this general discussion can be relevant to a physical situation, we focus on a work extraction protocol performed by an agent that is non-neutral to risk. We aim to obtain the fluctuation theorem in Eq.~\eqref{eq.fluc}, where  the wealth $w$ is equal to the work extracted. The work extraction will be realized by performing a cyclic change of the parameters of a thermally isolated system, which is initially prepared in a certain nonequilibrium state. To be more specific, this situation can be realized by focusing on a closed quantum system made of two parties $A$ and $B$ having Hamiltonian $H_A(t)$ and $H_B(t)$, which are time-dependent. By considering the time-dependent interaction $H_{int}(t)$ between $A$ and $B$, the total Hamiltonian $H(t)=H_A(t) + H_B(t) + H_{int}(t)$ will generate the unitary time evolution operator $U_{t,0}$, which is the solution of the Schr\"{o}dinger equation $i \dot U_{t,0}=H(t) U_{t,0}$ with initial condition $U_{0,0}=\mathds{1}$. In particular we consider the case where the interaction $H_{int}(t)$ is turned-off at the initial time $t=0$ and final time $t=\tau$. The work extraction is performed by cyclically changing some parameters of the system, so that $H(0)=H(\tau)=H$ and so $H_A(0)=H_A(\tau)=H_A$ and $H_B(0)=H_B(\tau)=H_B$. The initial state is locally at equilibrium, i.e.,  $\rho(0)= e^{-\beta_A H_A}\otimes e^{- \beta_B H_B}/Z$ where $Z$ is the normalization constant $Z =Z_A Z_B$, where $Z_X=\Tr{e^{-\beta_X H_X}}$ with $X=A,B$. Since the system is thermally isolated, the work extracted is minus the change of energy, so that the average work extracted is $\langle w \rangle = \Tr{H(\rho(0)-\rho(\tau))}$, where $\rho(\tau) = U_{t,0} \rho(0)U^\dagger_{t,0}$.
We note that if $\beta_A=\beta_B$ the initial state is a Gibbs state so that it is passive~\cite{allahverdyan04}, i.e., it is not possible to extract a non zero average work by performing any unitary cycle.
The work $w$ is defined by adopting a two-projective measurement scheme~\cite{talkner07}. Initially, we perform an energy measurement of the two parties $A$ and $B$ at the time $t=0$. Then we extract the work by changing the Hamiltonian in the time interval $(0,\tau)$ generating the time evolution with unitary operator $U_{t,0}$. In the end, we again perform an energy measurement of the two parties $A$ and $B$ at the time $t=\tau$. With the aim to get the fluctuation theorem in Eq.~\eqref{eq.fluc}, we define a stochastic entropy $\sigma$ with the joint probability distribution
\begin{eqnarray}
\nonumber p(\sigma,w) &=& \sum p_{m m'}P_{m m' n n'}\delta(\sigma-\sigma_{m m' n n'})\\
 && \times\delta(w-\epsilon^A_m-\epsilon^B_{m'}+ \epsilon^A_n+\epsilon^B_{n'})\,,
\end{eqnarray}
where the initial populations are $p_{m m'}= \bra{\epsilon^A_m,\epsilon^B_{m'}}\rho(0) \ket{\epsilon^A_m,\epsilon^B_{m'}}$, $\ket{\epsilon^X_m}$ are eigenstates of $H_X$ with eigenvalues $\epsilon^X_m$, with $X=A,B$, and $\sigma_{m m' n n'}=\beta_A(\epsilon^A_n-\epsilon^A_m)+\beta_B(\epsilon^B_{n'}-\epsilon^B_{m'})$. The transition probability is given by $P_{m m' n n'}=\abs{\bra{\epsilon^A_n,\epsilon^B_{n'}}U_{\tau,0} \ket{\epsilon^A_m,\epsilon^B_{m'}}}^2$.
\begin{figure}
[t!]
\centering
\includegraphics[width=0.87\columnwidth]{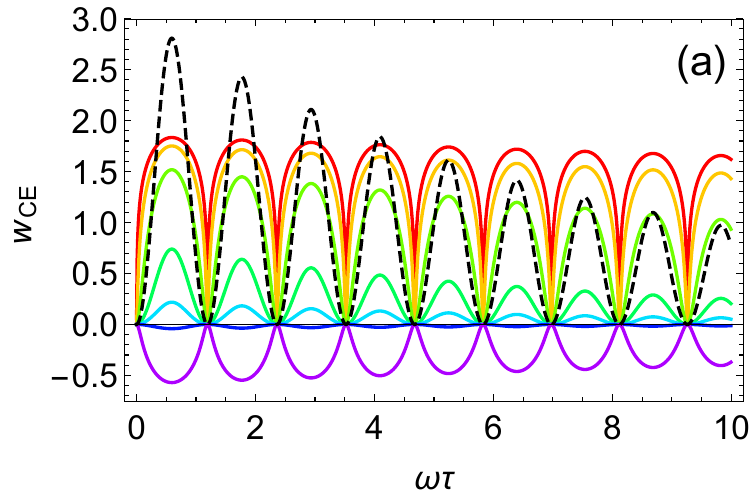}\\
\includegraphics[width=0.87\columnwidth]{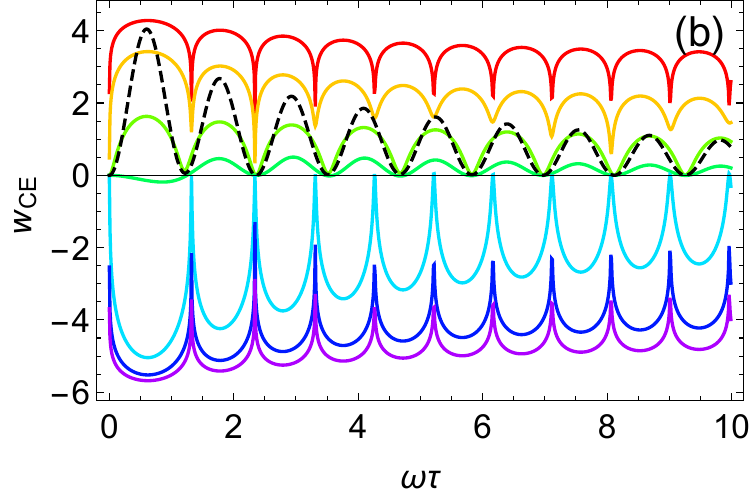}
\caption{ The plots of the average entropy $\langle\sigma\rangle$ (dashed lines), and the  certainty equivalents (solid lines) in the function of the duration time $\tau$. We consider $\omega=\min(\omega_A,\omega_B)/2$, $\lambda(t) = 2\omega \sin(\pi t/\tau)$, $\beta_A=0.1$, $\beta_B=4$, $\omega_A=2\omega_B$, $\gamma=0$ in the panel (a) and $\gamma=0.5$ in the panel (b). We put $r=-6,-4,-2,0,2,4,6$ from top to bottom.
}
\label{fig:plotfin3}
\end{figure}
It is easy to see that this probability distribution satisfies the fluctuation theorem in Eq.~\eqref{eq.fluc}.
To fix ideas, we focus on two qubits with local Hamiltonian $H_X(t)=\omega_X \sigma^X_z/2$, with $X=A,B$, with the interaction
\begin{equation}
H_{int}(t) = \lambda(t) ((1+\gamma)\sigma^A_x \otimes \sigma^B_x+(1-\gamma)\sigma^A_y \otimes \sigma^B_y)\,,
\end{equation}
where the coupling $\lambda(t)$ is such that $\lambda(0)=\lambda(\tau)=0$ and $\{\sigma_x,\sigma_y,\sigma_z\}$ are the Pauli matrices. The local ground states are $\ket{\epsilon^X_1}$ with energy $\epsilon^X_1=-\omega_X/2$, with $X=A,B$, while $\ket{\epsilon^X_2}$ are the local excited states with energy $\epsilon^X_2=\omega_X/2$. We consider $\omega_A>\omega_B$, it is easy to see that the initial state is non-passive if $\beta_B/\beta_A>\omega_B/\omega_A$, so that we expect $\langle w \rangle >0$ at least for certain unitary evolutions.
For $\gamma=0$ the excitation number is conserved and we can have only transitions in the sector with odd parity excitation, i.e., between the states $\ket{\epsilon^A_1 \epsilon^B_2}$ and $\ket{\epsilon^A_2 \epsilon^B_1}$. In this case the support of the distribution probability  of work has only three points, which are $w=\omega_B-\omega_A,0,\omega_A-\omega_B$. In contrast, for $\gamma\neq 0$, we can have also transitions in the sector with even parity excitation, i.e., also between the states $\ket{\epsilon^A_1 \epsilon^B_1}$ and $\ket{\epsilon^A_2 \epsilon^B_2}$, so that the support of the distribution probability of work has five points, which are $w=-\omega_A-\omega_B,\omega_B-\omega_A,0,\omega_A-\omega_B,\omega_A+\omega_B$.
To characterize risk aversion, we consider the exponential utility function $u(w)$ in Eq.~\eqref{eq.uteqxpo}.
As shown in Fig.~\ref{fig:plotfin3}, for $\gamma=0$ the imprint of the entropy is strong, which determines completely the trend of the certainty equivalent. However, for $\gamma\neq 0$, the certainty equivalent follows the trend of the entropy only for not too large $r$, otherwise $w_{CE}$ shows oscillations with different period with respect to the ones of $\langle \sigma\rangle$. This example shows how in general very risk averse (loving) agents cannot look to entropy alone to make their choice. On the other hand, the entropy can be a useful reference quantity for agents not too far from being risk neutral.

\section{Conclusions}

To summarize, we investigated what role fluctuation theorems can play in expected utility hypothesis. To achieve our results, we make the hypothesis that wealth is related to a stochastic entropy by a fluctuation theorem, differently from Ref.~\cite{Ducuara23} where a Crooks fluctuation theorem for the work is assumed instead. We prove the existence of a general bound for the certainty equivalent, which depends on the entropy, reproducing the thermodynamic uncertainty relation for a risk neutral agent. Thus, we examine the predictions and limitations of that bound.
Our aim is to answer the question of whether and when entropy can be a useful quantity to help a risk non-neutral agent make its decision.
It results that the relation between certainty equivalent and entropy can be simple as suggested by the bound found, but can also become complex as showed with some examples.
Finally, we apply our results to thermodynamics by focusing on a work extraction protocol realized by performing a unitary cycle which starts from an initial nonequilibrium state.

\subsection*{Acknowledgements}
The authors acknowledge financial support from the project BIRD 2021 "Correlations, dynamics and topology in long-range quantum systems" of the Department of Physics and Astronomy, University of Padova and from the European Union-Next Generation EU within the National Center for HPC, Big Data and Quantum Computing (Project No. CN00000013, CN1 Spoke 10 Quantum Computing).

\end{document}